# Coherent interaction of $WS_2$ and quasi-2D perovskite excitons over micrometer distances via a cavity field

*Marti Struve[1], Hamid Pashaei Adl[1], Jamie M. Fitzgerald[2], Oliwia Janikowska[3], Maciej Śmiertka[3], Alessandro Surrente[3], Sven Stephan[4], Christoph Lienau[1], Falk Eilenberger[5], Zdeněk Sofer[6], Watcharaphol Paritmongkol[7,8], William A. Tisdale[7], Paulina Plochocka[3,9], Ermin Malic[2], Christian Schneider[1] and Martin Esmann[1*]*

[1] Institut für Physik, Fakultät V, Carl von Ossietzky Universität Oldenburg, 26129 Oldenburg, Germany

[2] Department of Physics, Philipps-Universität Marburg, 35032 Marburg, Germany

[3] Department of Experimental Physics, Faculty of Fundamental Problems of Technology, Wroclaw University of Science and Technology, Wroclaw, 50-370 Poland

[4] University of Applied Sciences Emden/Leer, 26723 Emden, Germany

[5] Fraunhofer-Institute for Applied Optics and Precision Engineering IOF, Jena, Germany Institute of Applied Physics, Abbe Center of Photonics, Friedrich Schiller University, Jena, Germany

[6] Department of Inorganic Chemistry, University of Chemistry and Technology Prague, Technická 5, 166 28 Prague 6, Czech Republic

[7] Department of Chemical Engineering, Massachusetts Institute of Technology, Cambridge, MA 02139, USA

[8] Department of Materials Science and Engineering, School of Molecular Science and Engineering, Vidyasirimedhi Institute of Science and Technology (VISTEC), Rayong 21210, Thailand

[9] Laboratoire National des Champs Magnétiques Intenses, EMFL, CNRS UPR, 3228, University Grenoble Alpes, University Toulouse, University Toulouse 3, INSA-T Grenoble 38042 and, Toulouse, 31400 France

The coherent coupling of cavity-confined photons and excitonic matter resonances leads to the formation of cavity polaritons, hybrid light-matter quasi-particles. If multiple exciton resonances couple to the same photonic mode, the resulting polariton constitutes a coherent interaction between matter resonances that can be spatially separated without any direct electronic coupling. In this work, we demonstrate the formation of such a coherent coupling at room temperature using an open optical cavity containing two distinct van der Waals materials - monolayer $WS_2$ and layered quasi-2D halide perovskites (HaPs) - separated by 1.5 µm. The system forms three polariton branches, with the middle branch possessing nearly equal fractions of both excitons and the photonic mode. White-light reflectivity and luminescence measurements are in good agreement with simulations using a coupled harmonic oscillator and a microscopic Wannier-Hopfield framework. Our results lay the foundation to combine highly complementary degrees of freedom in 2D materials in an in-situ tunable fashion to enable new polaritonic functionalities.

## INTRODUCTION

Cavity-exciton polaritons are hybrid light-matter quasi-particles which inherit properties from all their constituents, such as the extended wavefunction of the photon, and polariton-polariton interactions triggered by their excitonic component. Van der Waals bound 2D materials are particularly promising for cavity-polariton devices due to their ease of fabrication and flexible hetero-integration. Furthermore, the large exciton binding energies make them room-temperature compatible. In particular, two types of materials stand out in the field: layered quasi-2D halide perovskites (HaPs) and transition metal dichalcogenides (TMDs).

Quasi-2D HaPs are chemically synthesized realizations of multi-quantum well (QW) stacks [1–6], hosting tightly bound room-temperature stable excitons ($E_b$~100 meV) [4,7]. Just as in traditional III-V multi-QW stacks, the number of layers scales up the overall exciton oscillator strength in quasi-2D HaP crystals. Choosing the thickness of the inorganic layers in terms of the number of unit cells, n, provides a unique control over the exciton resonances in quasi-2D HaPs [3,8–10]. Even the exciton binding energy is a separate, directly accessible tuning parameter in quasi-2D HaPs, e.g. via doping the organic spacers that act as electronic potential barriers between the inorganic QWs [11]. Most importantly, quasi-2D HaPs can be micromechanically cleaved and re-assembled in a well-defined manner [12–16]. This has already led to various demonstrations of polariton formation [7,17–23] and condensation [24,25].

Similarly, transition metal dichalcogenide monolayers are atomically thin, direct band gap semiconductors with tightly bound excitons that feature giant light-matter interactions [26–29]. These properties have enabled the observation of exciton-polaritons from cryogenic up to room temperature [30–33], polariton lasing and condensation, as well as polaritons subjected to photonic potential landscapes featuring non-trivial topology [34].

By hybridizing multiple excitonic resonators with a single photonic cavity mode, the resulting polariton inherits mixed properties from all contributing constituents. This concept has been implemented with all inorganic quantum wells [35] combinations of TMD and organic dyes[36], organic molecules and III-V-semiconductors [37], organic dyes and perovskites [38] or all organic molecular system[39,40]. In a broader sense, similar ideas have been applied to plasmon-exciton polaritons [41–45]. These systems have been put forward as promising platforms to address challenges such as understanding energy transfer processes via intermediate states [38,46] or generating new functionality in integrated opto-electronics [36,41].

The combination of quasi-2D HaPs and TMDs represents an ideal material platform for the realization of hybrid polariton devices, as their excitonic resonances can be brought into close energetic proximity, enabling hybridization, yet remains largely unexplored. In particular, it combines the large oscillator strength of the quasi-2D HaP with wide tunability of the exciton resonance energies via the perovskite quantum well thickness, the organic linkers, or the TMD. In addition, it presents an exciting outlook by combining a GaAs-type band structure and a soft-lattice phonon environment of the HaP with the valley degree of freedom of TMD monolayers, which can be optically accessed and controlled [30,32,47,48].

Here, we demonstrate the coherent hybridization between exciton resonances in a mechanically exfoliated layer of $WS_2$ and $(BA)_2(MA)_{n-1}Pb_nI_{3n+1}$ (n=4, BA= butylammonium, MA=methylammonium) quasi-2D HaP crystals. Despite their spatial separation of 1.5 µm, the coupling to one single photonic cavity mode in a tunable open cavity [33,49] at ambient conditions leads to the formations of three hybrid polaritonic branches, where the middle branch contains nearly equal contributions of photonic and excitonic fractions. We investigate cavity tuning-dependent reflectivity and luminescence properties of the resulting polaritons, finding convincing agreement with coupled oscillator and microscopic Wannier-Hopfield models. The tunable exciton energy and oscillator strength of the perovskite in conjunction with the in-situ tunable open cavity allow us to explore a large range of Hopfield coefficients and coupling strengths [7]. Specifically, the material combination explored in our work holds promise for new optoelectronic functionalities via the controllable mixing of distinct excitons, leveraging their respective large oscillator strengths and valley polarization. Unlike conventional van der Waals heterostructures, where interlayer contact leads to incoherent Förster energy transfer and ultrafast charge transfer that quench the individual excitonic resonances, the micrometer-scale separation in our architecture completely suppresses these short-range interactions. Consequently, the two excitonic systems interact exclusively through the shared cavity photon field, enabling coherent photon-mediated hybridization while preserving the intrinsic optical properties of each material. Furthermore, the interaction strength is externally tunable through cavity detuning, providing a level of control that is decoupled from the constituent material properties.

## RESULTS AND DISCUSSION

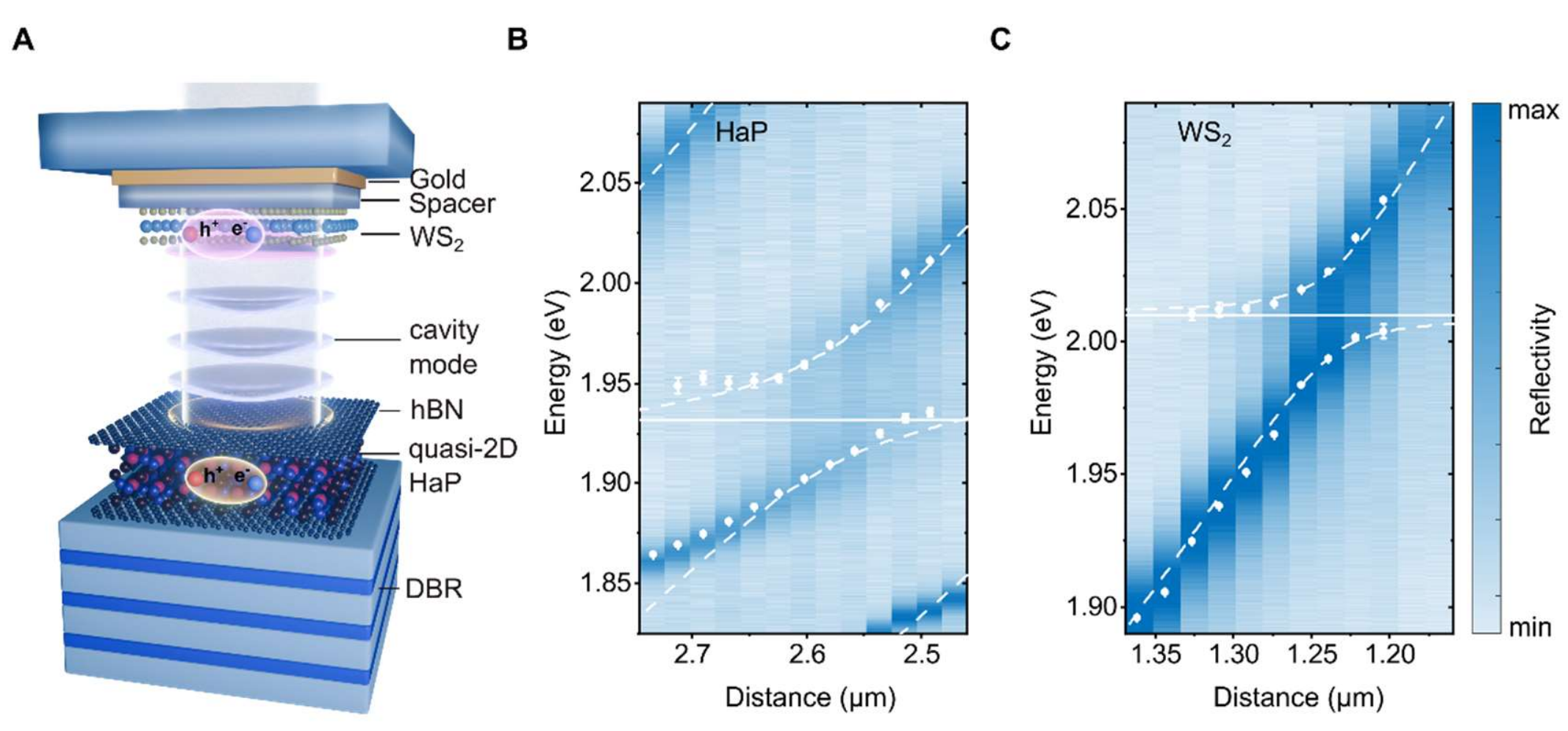

*Figure 1: Open cavity hosting a TMD and quasi-2D HaP and individual demonstration of the strong coupling conditions for both materials. A) An open cavity formed by one distributed Bragg reflector (DBR) and a gold mirror. Separate nanopositioner stages allow for precise control of the air gap and x-y-orientation. The bottom mirror contains a fully hBN-encapsulated 110-nm-thick n=4 quasi-2D halide perovskite (HaP) flake. To lift the $WS_2$ monolayer -crystal into a field maximum in front of the gold-coated top mirror, a 45-nm-thick hBN spacer is used. B) and C) White-light reflectivity at normal incidence as a functionof the cavity air gap for the individual quasi-2D halide perovskite (B) and the $WS_2$ monolayer crystal (C) respectively. The reflection minima are extracted via a Lorentzian-Fit (white dots). The emergence of a pronounced Rabi splitting around the respective exciton energies (solid white lines) demonstrates the formation of a cavity exciton-polariton in both cases. A coupled oscillator model is superimposed to the data (dashed white lines). From the model we extract a coupling strength of $g_{HaP}$=30 meV for the HaP flake and $g_{WS2}$ =16 meV for the $WS_2$ monolayer crystal, respectively.*

Fig. 1A shows the structure of the fully assembled experimental setup. The bottom mirror consists of 10 alternating quarter-wave pairs of $SiO_2$ and $TiO_2$, terminated with an additional $SiO_2$ layer, forming a distributed Bragg reflector (DBR). We transferred a fully hBN encapsulated ~110-nm-thick quasi-2D HaP crystal onto the DBR via dry stamping. As a top mirror we used a mesa structure laser-cut from a $SiO_2$ substrate that is coated with a 50-nm-thick gold film. Onto this gold mirror a $WS_2$-monolayer on top of a 40-nm-thick hBN flake is transferred. The hBN acts as a spacer to position the $WS_2$ into a field maximum of the cavity (see Supplementary Sections S2 for optical images of the sample, and S6 for further details on experimental methods). Each mirror is mounted on a separate piezo stack with xyz degree of freedom forming a fully flexible air-gapped Fabry-Pérot resonator [49].

To confirm strong coupling in our system, we performed white-light reflectivity (WL) measurements on both materials separately. First, we investigated the coupling behavior of the quasi-2D HaP crystal. The bottom mirror was adjusted so the HaP crystal was illuminated by the white-light source, while the top mirror was positioned over an area not covered by the hBN- $WS_2$-monolayer stack (Fig. 1B). Subsequently, the quasi-2D HaP was translated laterally and the hBN monolayer stack was positioned into the illumination area for independent characterization (Fig. 1C). In both configurations we systematically varied the cavity air gap and recorded angle-resolved WL spectra. The results from distance-dependent cross-sections through these spectra at $k_{||} = 0\ \mu m^{-1}$ are shown in Fig. 1B for the single quasi-2D HaP crystal and in Fig. 1C for the $WS_2$ monolayer. The reflectivity minima were extracted by fitting Lorentzian line shapes to each reflectivity spectrum in the distance series (white symbols in panels B-C). Through these minima we superimpose a coupled harmonic oscillator-model (dashed white lines). In the case of the $WS_2$ monolayer, an effective 2x2 Hamiltonian, consisting of a single exciton at $E^X_{WS2}$=2.001 eV and photonic mode $E^{C1}$, is sufficient to accurately describe the anti-crossing in Fig. 1C. In contrast, for the quasi-2D HaP shown in Fig. 1B, the larger light-matter coupling strength becomes comparable to the free spectral range of the empty cavity. This necessitates an effective 4x4 Hamiltonian, including one exciton at $E^X_{HaP}$=1.936 eV (see Supplementary Section S3) and three photonic modes $E^{C1}, E^{C2}, E^{C3}$.

At an air gap of 2.6 µm, calibrated by comparison with transfer matrix simulations, we extract a coupling strength of $g_{\mathrm{HaP}} = 30$ meV for the quasi-2D HaP flake. For the $WS_2$ monolayer, we extract a coupling strength of $g_{WS2}$=16 meV at an air gap of 1.25 µm in agreement with past studies [49]. To further substantiate the quantitative evidence for coherent coupling, we extracted the bare exciton linewidths from uncoupled reference measurements $WS_2$ flakes outside the cavity, yielding $\gamma_X^{WS2}$ = 34.3 meV. The empty cavity yielded a simulated Lorentzian Linewidth of $\gamma_C$ = 5 meV. We then explicitly evaluated the strong-coupling criterion, (g > ($\gamma_C$ + $\gamma_X$)/4)[50,51] for the excitonic resonance. The extracted coupling strengths $g_{WS2}$ = 16 meV exceed the corresponding threshold value of 9.75 meV, confirming that the criterion is satisfied.

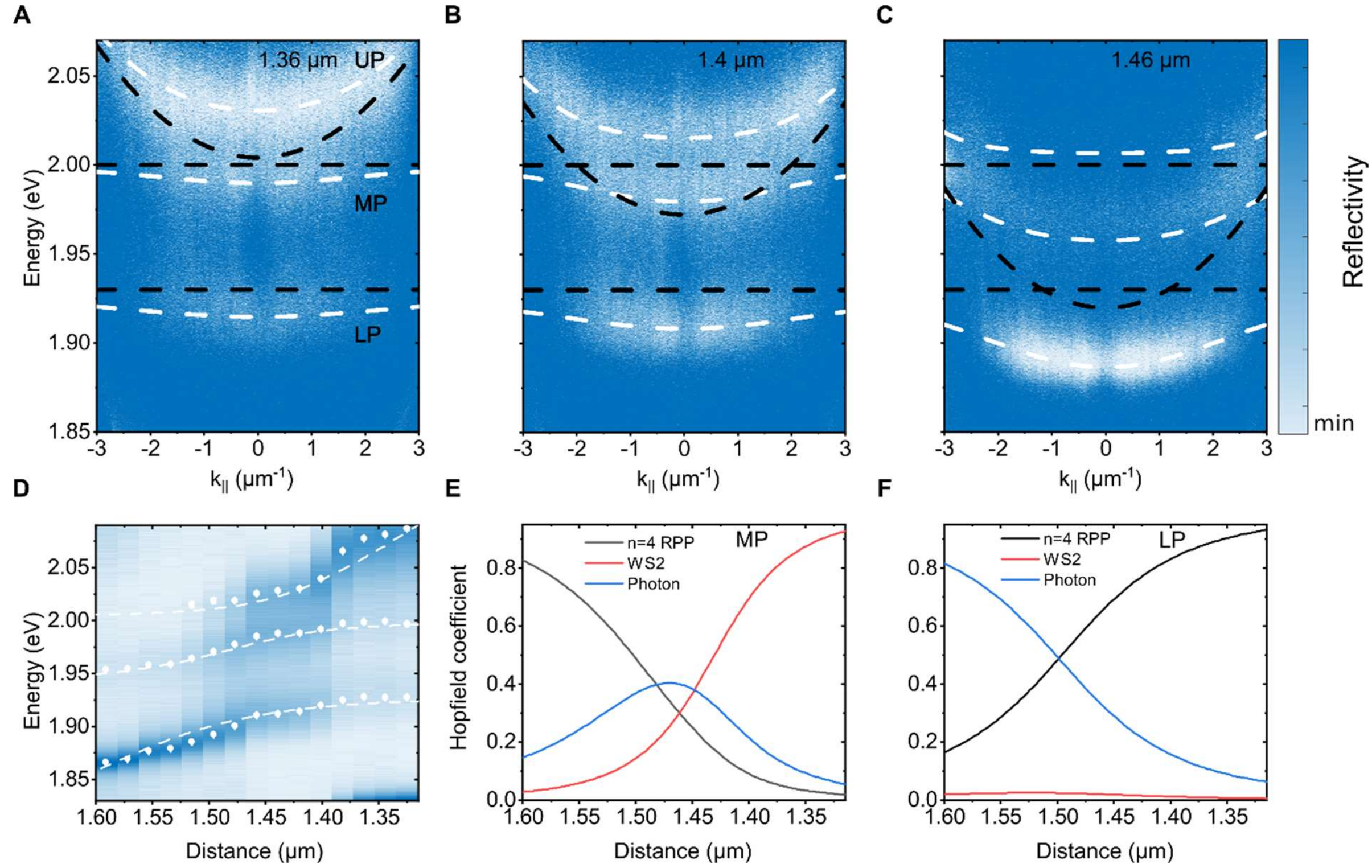

*Figure 2: Strong coupling and Hopfield coefficients in the hybrid TMD and quasi-2D HaP structure. A) – C) Angle-resolved white-light reflectivity spectra of the quasi-2D HaP flake together with the $WS_2$ monolayer crystal for three selected air gap sizes of 1.36 µm (A), 1.4 µm (B) and 1.46 µm (C), indicated at the top of each panel. A single cavity resonance couples to the two excitons (marked with black dotted lines) leading to the development of a lower (LP), a middle (MP) and an upper polariton (UP) branch. The reflectivity spectra are overlaid with a coupled oscillator model (dashed white lines) with coupling strengths $g_{HaP}$= 36 meV and $g_{WS2}$=20 meV. D) Air gap-dependent white-light reflectivity spectra at normal incidence extracted from the full distance-dependent series of angle-resolved WL spectra. Extracted reflection minima (white dots) are overlaid with a coupled oscillator model (dashed lines). E)-F) Hopfield coefficients extracted from the coupled oscillator model for the middle (E) and lower polariton mode (F). At the inversion point at 1.46 µm, the middle polariton shows a nearly equal tri-partite mixture of all constituents. At a cavity distance of 1.51 µm, the lower polariton shows a 2.5% contribution from the $WS_2$ -layer, 56.5% from the photon, and 41 % from the HaP, respectively.*

In the next step, we confirm the coherent hybridization of both excitonic resonances with a cavity mode in the full heterostructure stack. Using the in-situ tunability of the open cavity, the top and bottom mirror are positioned such that the quasi-2D HaP crystal and the $WS_2$ monolayer are at the same lateral position and both are illuminated by the white light simultaneously. Exemplary results of the angle-resolved reflectivity measurements are shown for air gaps of 1.36 µm, 1.4 µm and 1.46 µm in Fig. 2A-C. The spectra clearly reveal the emergence of three distinct dispersive branches. To model this behavior, we use a 3x3 coupled oscillator model consisting of one photon mode and two excitonic resonances at $E^X_{WS2}$= 2.001 eV and $E^X_{HaP}$= 1.936 eV described by the Hamiltonian

$$H = \begin{bmatrix} E^X_{WS2} & 0 & g_{WS2} \\ 0 & E^X_{HaP} & g_{HaP} \\ g_{WS2} & g_{HaP} & E^C \end{bmatrix} \quad (1)$$

From this model we extract coupling strengths of $g_{WS2}$= 20 meV and $g_{HaP}$ = 36 meV for the $WS_2$ and the quasi-2D HaP crystal, respectively. It is important to note that the larger coupling strengths extracted for the hybrid configuration arise from changes in the effective cavity geometry. For the perovskite flake, the reference measurements were performed in a cavity with an air-gap length of 2.6 µm, whereas the hybrid device was measured at an air gap of 1.36 µm. The resulting reduction in the effective cavity length increases the coupling strength. Likewise, the larger free spectral range allows the reduction to a 3x3 Hamiltonian in contrast to 4x4 in Fig. 1B. Fig. 2D shows the air gap-dependent angular cross-section of these spectra together with the extracted reflectivity minima. Using the same 3x3 coupled oscillator model, the experimental dispersion is reproduced (white dashed lines) describing the emergence of the upper (UP), middle (MP) and lower (LP) polariton branches (see Supplementary Section S4 for further transfer matrix simulations of the hybrid exciton-polariton system).

Based on this model we extract the Hopfield coefficients for the MP (Fig. 2E) and the LP (Fig. 2F). At a cavity distance of 1.46 µm our model predicts a MP with nearly equal tri-partite mixture of 40 % cavity mode and 30 % $WS_2$ and quasi-2D HaP exciton, respectively, i.e. the full hybridization of the three components. These Hopfield coefficients are in-situ tunable via the open cavity. For example at 1.51 µm the MP features 21.4% cavity mode, 9% HaP, 69.5% $WS_2$ At the same air gap of 1.51 µm, the LP consists of 2.5 % $WS_2$ contribution, 56.5% photon contribution and a 41 % HaP contribution (at 1.46 µm we have 33.4% photon, 64.6% HaP, 2% $WS_2$ for the LP).. These results clearly confirm the coherent coupling of two different exciton species across a micron-scale separation, mediated purely by strong light-matter interactions without direct electronic coupling.

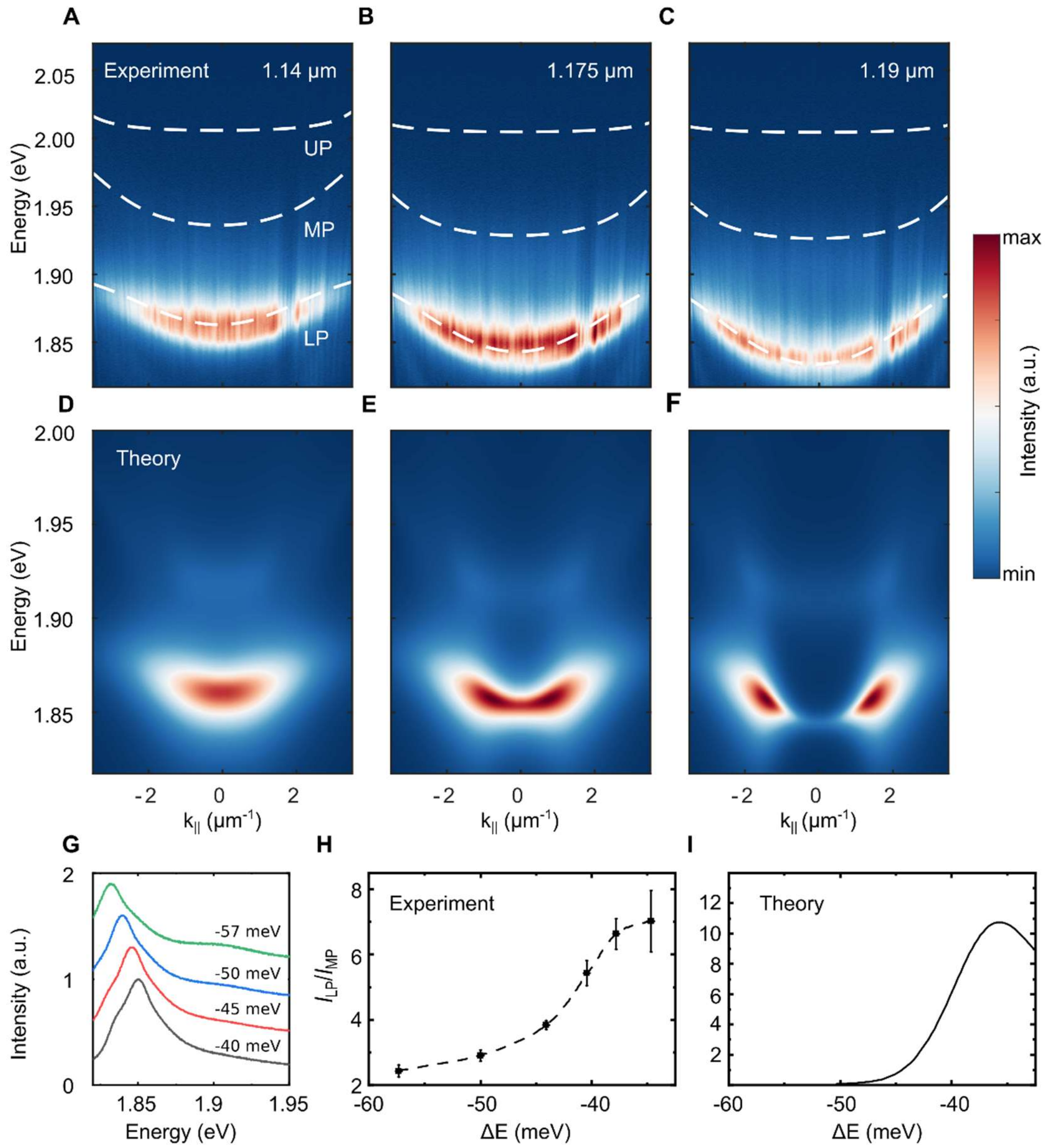

*Figure 3: Experimentally measured and simulated angle resolved photoluminescence. A) –C) Angle resolved photoluminescence (PL) spectra under 532 nm CW excitation for air gaps of 1.14 µm (A), 1.175 µm (B) and 1.19 µm (C), indicated at the top of each panel. The dashed lines show a coupled oscillator model superimposed with the spectra. At $k_{||}=2.5\ \mu m^{-1}$ the dispersion of the lower polariton branch features an inflection point. For strong red detuning (C) the PL intensity redistributes towards larger in-plane k-vectors indicative of a polariton bottleneck behavior. D)-F) Simulated angle-resolved PL using a microscopic Wannier-Hopfield model. An increase of the PL for higher in-plane k-vectors is observed for increasing red-detuning, in good qualitative agreement with the experimental data. G) PL spectra of the LP and MP branches extracted at $k_{||}=0\ \mu m^{-1}$ for different detunings between the LP and the HaP exciton. H) Experimentally extracted intensity ratio of the LP and MP at $k_{||}=0\ \mu m^{-1}$. I) Corresponding theoretically calculated LP-to-MP intensity ratio, reproducing the trend observed in the experimental data.*

Fig. 3A – C show angle-resolved photoluminescence (PL) spectra of the hybrid polariton. We excite the system with a green 532 nm continuous wave laser (excitation power 280 µW, spot size of 5 µm). Under these excitation conditions the we do not observe any photo induced degradation

of the perovskite. By tuning the open cavity air gap from 1.14 µm (Fig. 3A) to 1.19 µm (Fig. 3C) we study the dependence of PL on the Hopfield coefficients and exciton-photon detuning. A 3x3 coupled oscillator model is superimposed (dashed lines) indicating the dispersions of the UP, MP and LP branches. Regarding our primary experimental observations, we note two key features. First, the emission is dominated by the LP branch, while the MP and UP remain less intense over the investigated detuning range. Second, with increasing red detuning, the PL redistributes towards larger in-plane momenta, indicating reduced relaxation towards the polariton ground state and results in the bottleneck behavior.

We model the exciton polaritons using a microscopic Wannier-Hopfield framework [52,53] and experimentally extracted light-matter coupling parameters.

The simulation results are plotted in Fig. 3D - F showing good qualitative agreement with the experiment. For the small momenta and detuning values explored in this work, phonon scattering from the LP and MP branches into the momentum-dark excitons of the TMD (KΛ and KK' valleys) is negligible. Therefore, the phonon-driven dephasing of these branches is dominated by scattering into the perovskite exciton reservoir, i.e., optically dark excitons residing in high-momentum states outside the light cone. Polariton PL is described using the polariton Elliott formula [52]

$$I_Q(\hbar\omega) \propto (\hbar\omega)^2 \sum_{n=1}^{3} \frac{2\gamma_{nQ}\Gamma_{nQ}}{\left(E_{nQ}^{\mathrm{P}} - \hbar\omega\right)^2 + \left(\gamma_{nQ} + \alpha_{nQ} + \Gamma_{nQ}\right)^2} N_{nQ}^{\mathrm{boltz}} \tag{2}$$

where $n$ is the polariton branch index, $Q$ is the in-plane momentum, $N^{\mathrm{boltz}}$ is the Boltzmann distribution at room temperature, and $E^{\mathrm{P}}$ is the lower polariton energy. The total broadening comprises the polariton-phonon scattering rate $\Gamma$, the polariton radiative decay $\gamma$, and the absorptive loss $\alpha$ arising from the gold mirror. The radiative and absorptive terms correspond to the cavity decay and mirror loss rates, respectively, weighted by the photonic Hopfield coefficient squared. These bare cavity rates are extracted from a single-port coupled mode theory fit to an S-matrix calculation of the microcavity at each detuning[54]. Polariton-phonon scattering is treated via the deformation potential and second-order Born-Markov approximations [52,55]. Parameters for the n = 1 $(PEA)_2PbI_4$ structure are used [56], which is expected to provide an upper bound for the exciton-phonon coupling in the n = 4 system [57,58], while the phonon energy is expected to be consistent across different n [59,60]. Based on these assumptions, we expect the results to remain qualitatively valid for the experimental system. We consider an averaged optical phonon with energy $E^{\mathrm{ph}}$=35 meV, consistent with the bottleneck observed when the cavity is red detuned. To mimic the impact of disorder, we incorporate inhomogeneous broadening by averaging the PL over a Gaussian exciton energy distribution with a standard deviation of 5 meV. This lessens the severity of the polariton bottleneck, yielding closer agreement with experimental observations.

Equation 2 describes the compromise between a large polariton occupation and photonic character needed to maximize PL. Polariton states with energies below the perovskite exciton energy minus the optical phonon energy cannot be populated via direct phonon scattering and hence will thermalize to a depleted Boltzmann distribution, i.e., a polariton bottleneck [52]. By tuning the cavity length, the momenta at which this phonon scattering channel opens can be modified. At a cavity length of 1.15 µm, the energy separation between $Q = 0$ polaritons and the perovskite 1s exciton energy is small enough that they can be directly populated via optical phonon scattering from the exciton reservoir. As the cavity length is increased to 1.16 µm, the LP redshifts and the Boltzmann factor in Eq. 2 leads to an increase in PL intensity. Finally, at a cavity length of 1.17 μm, the energy separation between the low-momentum polaritons and the exciton reservoir is larger than $E^{\mathrm{ph}}$, leading to a bottleneck for momenta below ~ 1 μm$^{-1}$ in good qualitative agreement with the experimentally measured PL intensity.

Similar logic can be applied to understanding the ratio between the LP and MP peak PL intensity. Figure 3G shows PL spectra at $k_{||} = 0\ \mu m^{-1}$ for different detunings between the LP and the HaP exciton. From these data, we extract the intensity ratio $I_{\mathrm{LP}}/I_{\mathrm{MP}}$, which is shown in Fig. 3H as a function of the energy detuning between the LP and the exciton resonance $E^{\mathrm{P}}$. The ratio exhibits a pronounced maximum of approximately 7 at a detuning of −34.6 meV and decreases to 2.4 at −57 meV. As the cavity is further red-detuned, the optical phonon scattering channel into the perovskite exciton reservoir opens for the lower polariton, removing the bottleneck and leading to an increased occupation and hence PL (1.18 to 1.16 μm). The peak PL intensity is governed by the critical coupling condition $\Gamma_{nQ} = \gamma_{nQ} + \alpha_{nQ}$, which maximizes the prefactor in front of the Boltzmann distribution in Eq. 2 when evaluated at $\hbar\omega = E^{\mathrm{P}}_{nQ}$ . For the lower polariton this is reached when the optical phonon scattering pathway opens, close to 1.16 μm, leading to a sharp increase in $\Gamma_{1Q}$. For shorter cavity lengths (1.16 to 1.15 μm), the lower polariton PL weakens due to a decreasing photonic character ($\Gamma_{1Q}$ becomes much larger than $\gamma_{1Q} + \alpha_{1Q}$). In contrast, the absorption scattering channel between the middle polariton and the perovskite exciton reservoir is always open, but close to cavity lengths of 1.14 μm, the phonon emission channel opens (i.e., middle polaritons at $Q = 0$ can scatter into the perovskite reservoir via the emission of an optical phonon). In this case, the rapid rise in $\Gamma_{2Q}$ actually takes the middle polariton away from the critical coupling condition, leading to a sharp decrease in PL for cavity lengths shorter than 1.14 μm. This means the maximum middle polariton emission occurs at shorter cavity lengths, just before the opening of the emission channel, leading to the peak in $I_{\mathrm{LP}}/I_{\mathrm{MP}}$ observed in the experiment. The corresponding theoretical results for the PL ratios of the two branches in Fig. 3H reproduce the trend in the experimental data qualitatively, confirming this interplay between phonon-assisted scattering channels and absorption scattering.

## CONCLUSION

In this joint experiment-theory work, we demonstrated the coherent coupling of excitonic resonances in two different van der Waals materials across a 1.5 μm air gap. By placing a $WS_2$ monolayer and n=4 quasi-2D HaP crystals on the two mirrors of a length-tunable, room-temperature open cavity, we observed the formation and photoluminescence of hybrid cavity exciton-polaritons. Our choice of the energy difference between the bare excitonic resonances in conjunction with selection of the individual coupling strengths, allows us to tune the Hopfield-coefficients of the middle polariton across the whole available range [7]. This unprecedented level of flexibility in a hybrid polariton system opens an interesting avenue to new polariton-based opto-electronic applications. In particular, we showcased a system where large oscillator strength, valley degree of freedom in the TMD, and potentially electric tenability [36,61] are all easily integrated into one experimental platform.

## ACKNOWLEDGMENTS

The authors acknowledge support by the European CommissionERC Dual-Twist (Grant number 101170213) and the Niedersächsische Ministerium für Wissenschaft und Kultur Wissenschaftsraum ElLiKo

The Marburg group acknowledges financial support by the Deutsche Forschungsgemeinschaft (DFG) via the regular projects 504846924 and 524612380

Z.S. was supported by ERC-CZ program (project LL2101) from Ministry of Education Youth and Sports (MEYS) and by the Advanced Multiscale Materials for Key Enabling Technologies project, supported by the Ministry of Education, Youth, and Sports of the Czech Republic. Project No. CZ.02.01.01/00/22_008/0004558, Co-funded by the European Union.

W.A.T. and W.P. were supported by the U.S. Department of Energy, Office of Science, Basic Energy Sciences, under award number DE-SC0019345.

# Supplementary Information

## S1: Microscopic Wannier-Hopfield framework

Exciton binding energies and wavefunctions of an $n = 1$ $(PEA)_2PbI_4$ quantum well structure are calculated microscopically by solving the two-dimensional Wannier equation (2) within the two-band parabolic approximation where the reduced mass of the electron and hole is $0.108m_e$ [2]. The screened Coulomb potential is modelled with a Rytova–Keldysh potential [3], where the high-frequency dielectric constants of the organic spacer layer and the inorganic perovskite layer are set to 3.32 and 6.1 [4], respectively. The thickness of the perovskite slab is taken as 0.636 nm [4]. This yields a 1s exciton binding energy of 237 meV. The spectral position of the exciton is fixed to the experimental value of 1.89 eV. Similarly, the 2D Wannier equation for the hBN-encapsulated $WS_2$ monolayer was solved using DFT input for the two-band parabolic approximation of the electronic bandstructure [5]. The screened Coulomb interaction is modelled using a generalised Keldysh potential [6] using DFT-calculated dielectric constants for the monolayer ($\epsilon_\perp = 7.2$ and $\epsilon_\parallel = 16.8$) [7], and $\epsilon_{\text{sub}} = 4.5$ for the encapsulating hBN layers. The spectral position of the TMD 1s exciton is like-wise fixed to its experimental value of 2.0 eV. The bare cavity dispersion, $E^c$, is approximated as parabolic with the effective mass extracted at each detuning from S-matrix simulations. Cavity decay rates are also estimated by fitting the calculated reflection coefficient using a single-port coupled mode theory [8]

$$r(\hbar\omega) = r_d(\hbar\omega) + \frac{2\gamma^c e^{i2\Phi}}{i(E^c - \hbar\omega) + \alpha^c + \gamma^c}, \tag{1}$$

where $\gamma^c$ denotes the radiative decay rate of the bare cavity mode, $\alpha^c$ represents the absorptive loss from the gold mirror, and $\phi$ is the phase shift that light accumulates during the process of coupling from the port into the cavity mode. The term $r_d$ accounts for the direct, non-resonant reflection, and is determined using a separate S-matrix calculation of the gold mirror alone. The material dispersion of gold is taken from experimental data [9]. A single-port description is appropriate as the DBR mirror has a much higher reflectance than the gold mirror, resulting in negligible transmission through the DBR. The exciton-cavity photon coupling strengths for the perovskite, $g_{Pe} = 30$ meV, and WS2 monolayer, $g_W = 16$ meV, are extracted from the from the coupled oscillator fit of the individual reflectance spectra. The polariton energies, $E^P_{nQ}$, and Hopfield coefficients, $U_{nm,Q}$, are obtained by diagonalizing a three-coupled-oscillator Hamiltonian describing a single cavity photon coupled to two distinct exciton resonances:

$$H = \begin{bmatrix} E^c_Q & g_{\text{Pe}} & g_W \\ g_{Pe} & E^{\text{Pe}}_Q & 0 \\ g_W & 0 & E^W_Q \end{bmatrix}, \tag{2}$$

where n is the polariton branch index, $Q$ is the in-plane momentum magnitude, and $E^{\mathrm{Pe}}$ and $E^{\mathrm{W}}$ are the perovskite and $WS_2$ exciton energies, respectively.

Electron-phonon matrix elements are described within the deformation potential approximation using parameters extracted in [3] from experimentally measured temperature-dependent linewidths of the 1s exciton in the $n = 1$ $(PEA)_2PbI_4$ structure. For $WS_2$, DFT parameters from Ref.[10] are used. Converting to the polariton basis gives the polariton-phonon matrix elements, as described in previous studies [11–14]. They are given by the 1s exciton-phonon matrix element [15], $D_{uv,|\vec{Q}'-\vec{Q}|}$, scaled by the relevant excitonic Hopfield coefficients of the initial and final polariton states: $\widetilde{D}_{n\vec{Q},m\vec{Q}'} = \sum_{uv} U^{*}_{un,\vec{Q}}\, D_{uv,|\vec{Q}'-\vec{Q}|} U_{vm,\vec{Q}'}$ , where $u$ and $v$ are exciton indices. In hybrid polaritonic systems where the cavity mediates interactions between spatially separated materials, the polariton-phonon matrix element must be calculated using only the excitonic Hopfield coefficient of the material hosting the specific phonon. The polariton-phonon scattering rates are calculated within the Born-Markov and phonon-bath approximations [11,16],

$$\Gamma_{n\vec{Q}} = \frac{\pi}{\hbar} \sum_{m\vec{Q}',\nu,\sigma} \left|\widetilde{D}_{\nu,n\vec{Q},m\vec{Q}'}\right|^2 \left(\frac{1}{2} + \sigma\frac{1}{2} + n^{\mathrm{ph}}_{\nu,|\vec{Q}-\vec{Q}'|}\right) \delta\left(E^{\mathrm{P}}_{m\vec{Q}'} - E^{\mathrm{P}}_{n\vec{Q}} + \sigma E^{\mathrm{ph}}_{\nu,|\vec{Q}-\vec{Q}'|}\right) \tag{3}$$

where $n^{ph}$ is the Bose-Einstein distribution, and we sum over all energy- and momentum-conserving scattering channels from a polariton in state $|n,\vec{Q}\rangle$ to all possible final states $|m,\vec{Q}'\rangle$. All relevant phonon modes (here, an averaged acoustic and optical $\Gamma$-phonon for each material) are summed over, denoted by $\nu$, as well as phonon emission/absorption ($\sigma = \pm 1$). Long-range acoustic phonons are approximated with a linear dispersion, while optical phonons are treated as dispersionless [3]. In particular, the averaged optical phonon energy of the perovskite layer is taken as $E^{\mathrm{ph}} = 35$ meV [3]. To numerically evaluate the scattering integral, we apply a Lorentzian broadening (2 meV) to relax strict energy-momentum conservation [12,13]. The steady-state polariton photoluminescence (PL) is calculated assuming that phonon-driven scattering into the light cone is dominated by exciton reservoir states from outside the light cone [11], giving Eq. 2 of the main text. The polariton radiative decay is given by the total bare cavity decay (evaluated at $Q = 0$) scaled by the photonic Hopfield coefficient, $\gamma_{nQ} = |U_{0n,Q}|^2\gamma^c$. Likewise, the polariton absorptive loss rate due to the gold mirror is $\alpha_{nQ} = |U_{0n,Q}|^2\alpha^c$. The absorption loss of the cavity modifies the PL in two ways. First, it acts an additional contribution to non-radiative polariton decay, decreasing the quantum efficiency of the system. The PL can be written in terms of the polariton occupation, $N^{\mathrm{P}}_{nQ}$, as [11] :

$$I(\hbar\omega) \propto \sum_{n=1}^{3} \frac{2\gamma_{nQ}(\gamma_{nQ} + \alpha_{nQ} + \Gamma_{nQ})}{\left(E^{P}_{nQ} - \hbar\omega\right)^2 + \left(\gamma_{nQ} + \alpha_{nQ} + \Gamma_{nQ}\right)^2} N^{\mathrm{P}}_{nQ}\,, \tag{4}$$

Second, the occupation itself is modified as there is now an additional decay channel that removes polaritons from inside the light cone, but does not impact reservoir excitons outside [11]

$$N^{\mathrm{P}}_{nQ} = \frac{\Gamma_{nQ}}{\gamma_{nQ} + \alpha_{nQ} + \Gamma_{nQ}}\, N^{\mathrm{boltz}}_{nQ}. \tag{5}$$

Hence, if there is no polariton radiative or cavity loss, then the polariton occupation will follow a Boltzmann distribution. Combining Eqs. 4 and 5, along with a factor of $(\hbar\omega)^2$ to account for the free-space photon density of states, yields Eq. 2 of the main text. The resulting equation is consistent with the Kubo-Martin-Schwinger relation between emission and absorption at thermodynamic equilibrium [17].

## S2: Optical microscope images of the sample

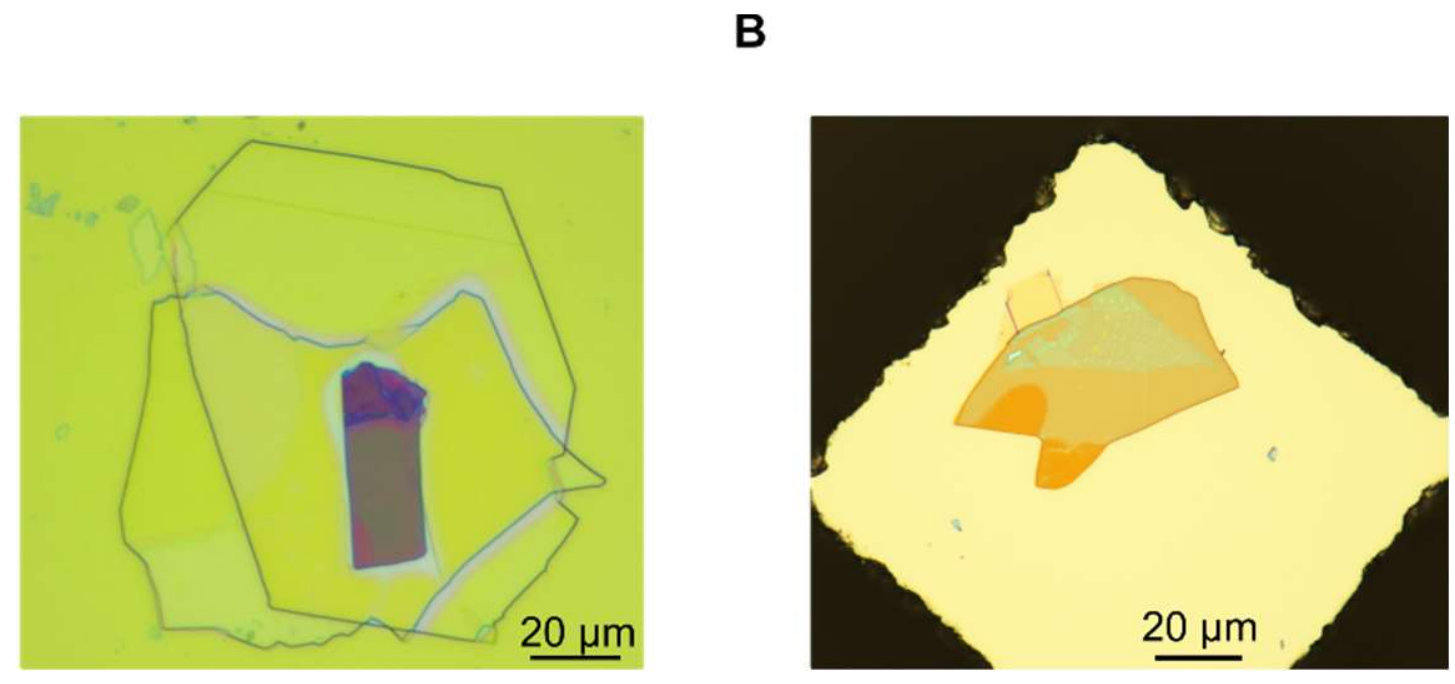

*Figure S1: **Optical microscope images. A)-B**) Optical microscope image of the fully encapsulated quasi-2D HaP flake (A) and the $WS_2$-monolayer on an hBN spacer (B). The encapsulated quasi-2D HaP flake is deposited on a DBR. The dark red region is the perovskite, while the transparent regions are the encapsulating hBN flakes. The $WS_2$-hBN stack is deposited on a gold-coated mesa-type structure (large bright square). The orange area is the 40-nm-thick hBN spacer that lifts the blueish $WS_2$ -monolayer into the field maximum in front of the metallic mirror.*

## S3: Photoluminescence of the quasi-2D HaP and $WS_2$ monolayer crystal

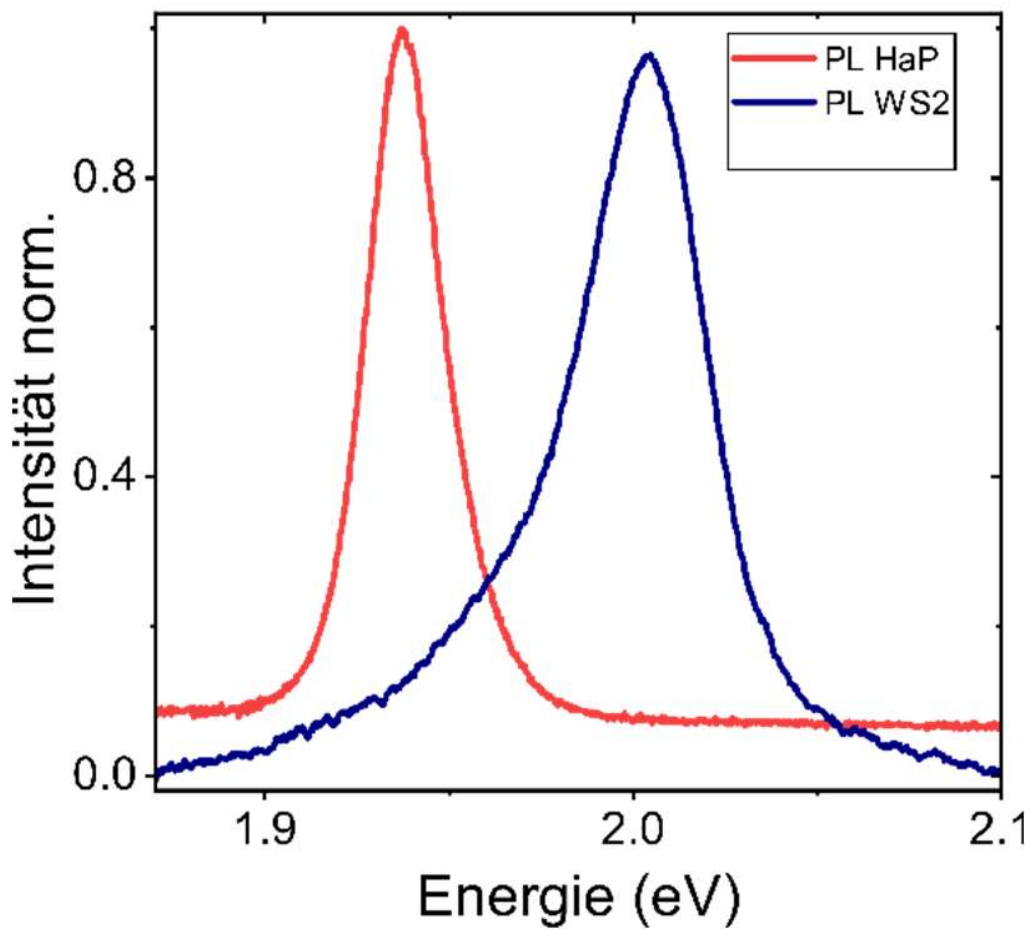

*Figure S2**: Photoluminescence spectra**. Photoluminescence spectra of a pure phase n=4 quasi-2D HaP flake (red) and a $WS_2$ -monolayer crystal (blue) under CW excitation at 532 nm.*

## S4: Transfer matrix simulation and comparison to experimental reflectivity

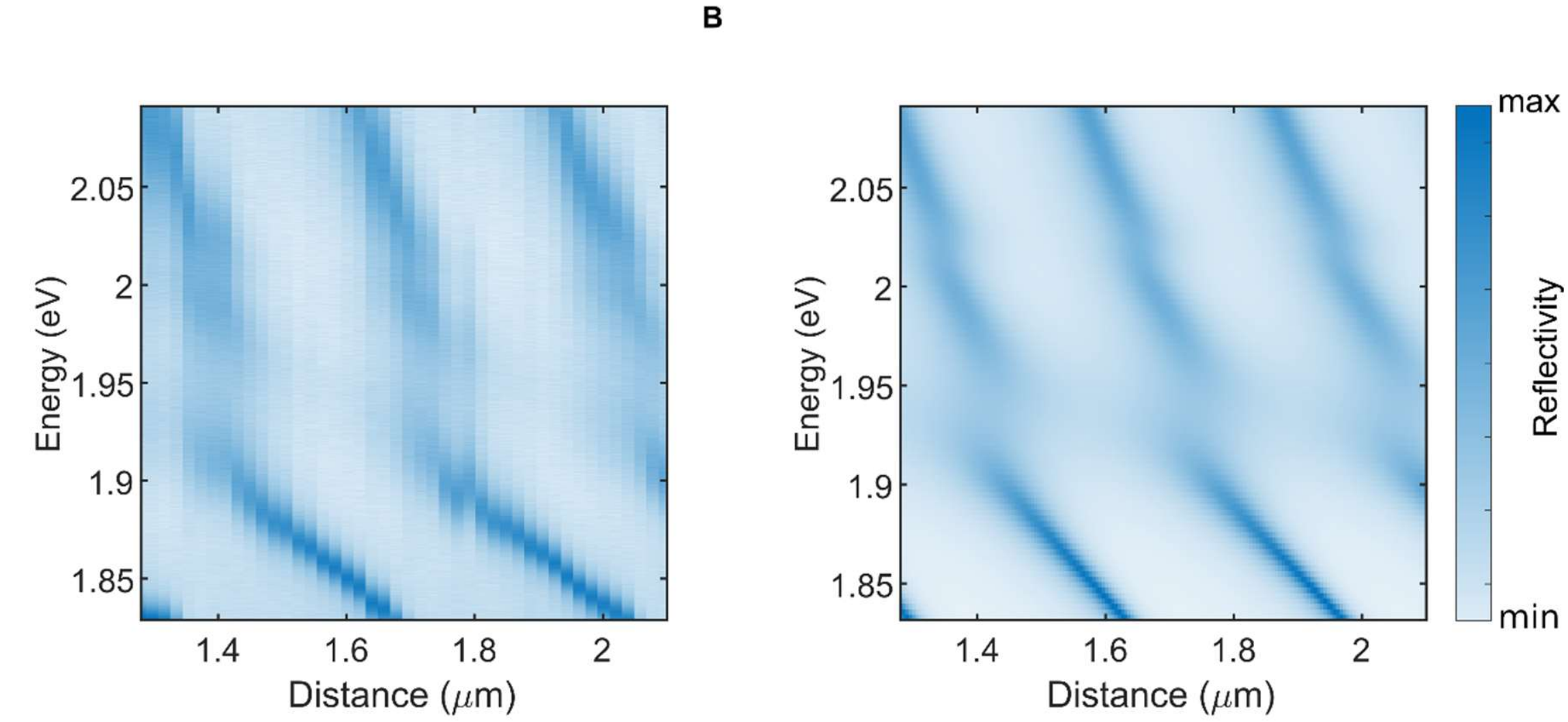

*Figure S3:* ***Cavity tuning series with extended range and corresponding transfer matrix simulation****.* ***A)*** *Experimental data of the distance series for the hybridized polariton shown in Fig. 2D of the main text with an extended distance axis.* ***B)*** *Corresponding transfer matrix simulation. Refractive index for the* $\mathbf{n = 4}$ *quasi-2D HaP was taken from Ref.* [18] *yielding good quantitative agreement with the experimental white light spectra.*

## S5: Optical Setup

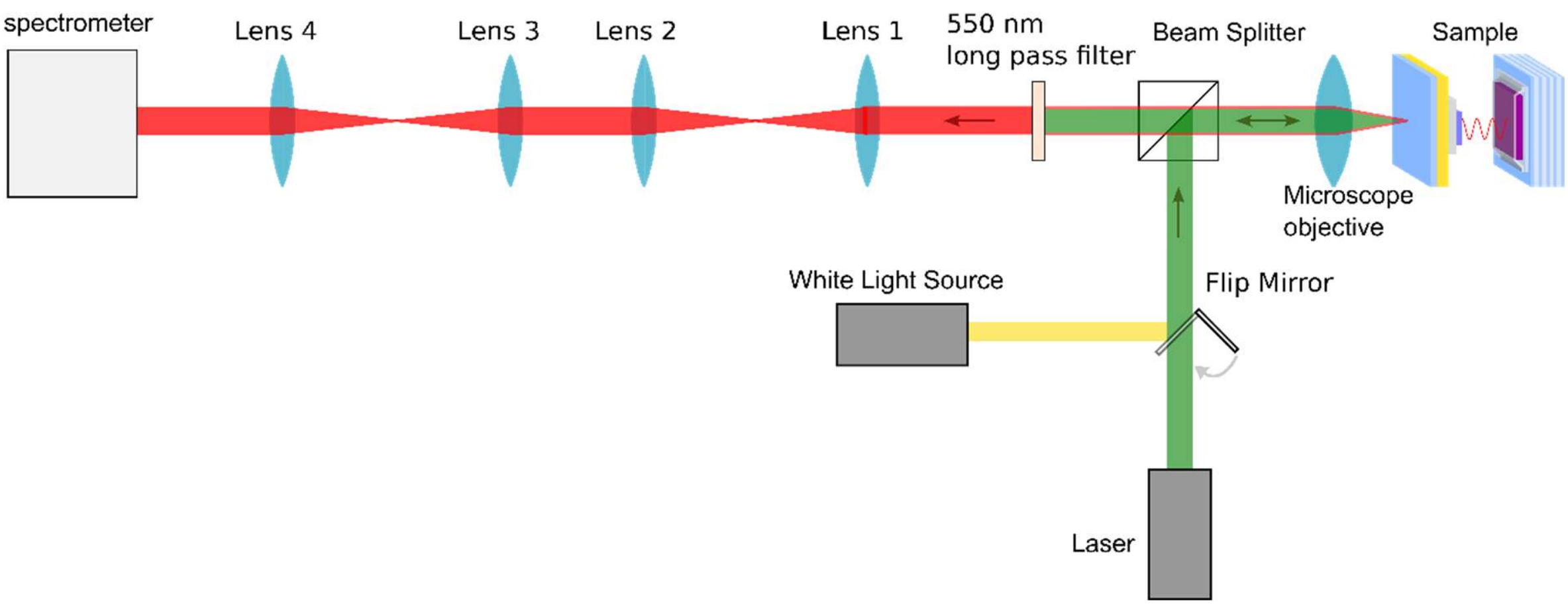

*Figure S4 :* ***Schematic of the optical setup****. Schematic of the Fourier-imaging reflectivity setup used to measure angle-resolved white light reflectivity (WL) and photoluminescence (PL).*

In our measurements, we use two different kinds of excitation that can be switched with a flip mounted mirror. For white-light reflectivity measurements, we use a white light source (Thorlabs SLS301), while for PL measurements we use a 532 nm continuous-wave DPSS laser. A 50:50 beam splitter directs the excitation beam into a microscope objective, which focuses it onto the sample. The active materials are mounted on two mirrors on separate piezo-motors with xyz degrees of freedom forming an open cavity as described in [19,20].

We collect the signal in a reflectivity configuration, using the microscope objective to collect the emission and transmit it through the beam splitter. To filter out any reflected excitation laser light, a 550 nm long pass filter is introduced in the beam path. Lens L1, positioned with its focal point on the back aperture of the microscope objective, directs the emission to a spatial filter, after which the beam is collimated by Lens L2. Lens L3 then images the back focal plane of the microscope objective. Through Lens L4, we image this back focal plane onto the monochromator (Andor Shamrock 500i) and to a Peltier-cooled EMCCD camera (Andor iXon Ultra 888). This configuration allows the collection of angle-resolved emission spectra.

**S6: Methods**

Sample Preparation

The bottom mirror is quarter-wave stack with 10 pairs of alternating layers of 107-nm-thick $SiO_2$ and 67-nm-thick $TiO_2$ terminated with an additional $SiO_2$ layer grown by HF sputtering. The top mirror consists of a 50-nm-thick gold film grown on a $SiO_2$-subtrate with a laser-cut 100 µm x 100µm x 100µm mesa-type structure using electron beam evaporation.

For the bottom mirror, all further fabrication steps were conducted in an inert nitrogen atmosphere inside a glovebox to avoid exposure to oxygen and moisture. A PPC/PDMS stamp was prepared by drop-casting a thin polypropylene carbonate (PPC) film onto a glass slide, allowing it to dry, and then transferring it onto a polydimethylsiloxane (PDMS) dome mounted on a glass slide, following the procedure described in [21]. All pick-up and transfer steps were performed at ~60 °C. Hexagonal boron nitride (hBN) flakes were mechanically exfoliated from bulk crystals (grown by the Chemical Vapor Transport) using Scotch tape and transferred onto a Si/$SiO_2$ substrate (90 nm oxide). A selected flake was then picked up using a stamp. $(BA)_2(MA)_{n-1}Pb_nI_{3n+1}$ (n = 4) perovskite crystals were similarly exfoliated from bulk crystals (grown by the cooling induced crystallisation) onto a separate substrate, and a ~110-nm-thick flake (estimated from optical contrast) was picked up by the hBN flake already attached to the PPC/PDMS stamp. Following this, a second hBN flake was picked up by the hBN/perovskite stack, resulting in the formation of an hBN/perovskite/hBN heterostructure [21]. The top and bottom hBN thicknesses were ~15-nm and ~20-nm, respectively, as determined by atomic force microscopy (AFM). The heterostructure was then aligned and transferred onto a distributed Bragg reflector (DBR). This final transfer step was performed at room temperature, during which the stack detached from the PPC/PDMS stamp upon contact with the substrate.

On the top mirror a hBN spacer and a $WS_2$ monolayer were deposited using the PDMS dry stamp method at atmospheric condition. First, the crystals were micromechanically cleaved at ambient conditions. We first transferred the 40-nm-thick hBN flake (2D Semiconductors, substrate at 60 °C, 5 min contact) onto the gold surface to act as a spacer that lifts the $WS_2$ monolayer (2D Semiconductors, substrate at 60 °C, 5 min contact), which was subsequently transferred in a second dry-stamping step into a cavity field maximum.

Data analysis

We extract the white-light reflectivity minima by fitting a Lorentzian function to the spectra. By using transfer-matrix-simulations, we reproduce the experimental data. From this, we calculate the uncoupled photon modes by neglecting the excitonic resonance and assuming only a constant real refractive index for the optically active material. The refractive indices for the HaP were taken from [18]. We couple these modes to the corresponding exciton with a coupled oscillator model to extract the coupling strength of the material.